\documentclass[12pt,preprint]{aastex}  

\begin{document}

\title{Predicting Ranges for Pulsars' Braking Indices\footnote{To appear in \textit{The Astrophysical Journal}, doi:10.1088/0004-637X/753/1/1.}}

\author{Nadja S. Magalhaes and Thaysa A. Miranda}

\email{nadjasm@gmail.com}

\affil{Federal University of Sao Paulo, DCET \\ DCET, Rua Sao Nicolau 210, Diadema, SP 09913-030, Brazil}

\and

\author{Carlos Frajuca}

\affil{Federal Institute of Education, Science and Technology of Sao Paulo \\ R.
Pedro Vicente 625, Sao Paulo, SP 01109-010, Brazil}

\begin{abstract}
The theoretical determination of braking indices of pulsars is still an open problem. In this paper we report results of a study concerning such determination based on a modification of the canonical model, which admits that pulsars are rotating magnetic dipoles, and on data from the seven pulsars with known braking indices. In order to test the modified model we predict ranges for the braking indices of other pulsars. 
\end{abstract}



\section{Introduction}

Pulsars are modeled as rapidly rotating, highly magnetized stars composed mainly of neutrons. It is observed that their rotation periods are increasing, implying a decay in their rotation frequencies. This spin down is quantified by the braking
index, $n$:
\begin{equation}
n  \equiv \frac{{\Omega \ddot \Omega }}
{{\dot \Omega ^2 }},\label{eq:def_n}
\end{equation}
where $\Omega$ is the pulsar's rotational angular velocity and the dot denotes a time derivative. In the canonical model the main time-varying field responsible for the loss of rotational energy in a pulsar is a magnetic dipole field \citep{osg}.

Several theoretical calculations have been tried to explain the observed values
for the braking index as can be found,  for instance, in \citet{blandRo88}, \citet{mela97} or \citet{contop06}. As far as the authors know, however, no theory was developed that satisfactorily explains all the values of the braking indices obtained from experimental data. In the cases of the pulsars
for which this index was measured it lies within the range $0.9 -
2.8$ (see Tables \ref{table:pulsars} and \ref{table:B}), while the canonical theory predicts $n=3$ in the approximation of magnetic dipole field dominance. Improvements on
this theory have been tried since pulsars were discovered, decades ago. It is known that several factors may affect the
braking index \citep{man77,kaspi94,glend96,lorimer08}.

In this paper we analyze a modification of the canonical model aiming at predicting braking indices of pulsars based on those that have already been measured. The modification consists basically in allowing for a variable braking index in the equation for the braking torque. This implies the introduction of a parameter that balances the units in the equation that is essentially related to the star's physical characteristics (mass, radius, magnetic field, etc.)

The prediction of braking indices will be tried using two measured parameters: the rotation frequency and its first time derivative. The reason for this choice is the possibility of predicting ranges of braking indices for a significant number of pulsars since many have these parameters measured already \citep{pulsarCatalog12}. The determination of exact values of braking indices demands more information, like the frequency's second time derivative or the angle between the rotation axis and the magnetic axis of the star, but these will not be considered in this work.

In the next section we present a summary of the canonical model and its modified version proposed by us. The following sections provide the results using the modified model, an analysis concerning them and concluding remarks.



\begin{deluxetable}{ccccc}
\tablecaption{\label{table:pulsars} Rotation frequencies ($\nu_0$) and their first time derivatives for the pulsars with known braking indices. Crab is PSR B0531+21 and Vela is PSR B0833$-$45.}
\tablewidth{0pt}
\tablehead{
  \colhead{Pulsar} & \colhead{$\nu_0$ (Hz)}  & \colhead{$\dot \nu_0$ ($\times 10^{-10} s^{-2}$)} & \colhead{$\ddot \nu_0$ ($\times 10^{-21} s^{-3}$)}  & \colhead{Refs} 
}
\startdata
  Crab &30.22543701 & $-$3.862283 & 12.4265 & 1 \\
  PSR B1509$-$58 & 6.633598804 & $-$0.675801754   & 1.95671 & 2   \\
  PSR B0540$-$69 & 19.8344965  & $-$1.88383 &   3.81 & 2, 3 \\
  PSR J1119$-$6127 & 2.4512027814 & $-$0.2415507 & 0.6389 & 4 \\
  Vela & 11.2 & $-$ 0.157 & 0.031 & 5 \\
  PSR J1846$-$0258 & 3.0621185502 & $-$0.6664350 & 3.13 & 6 \\
  PSR J1734$-$3333 & 0.855182765 & $-$0.0116702 & 0.0028 & 7 \\
\enddata
\tablerefs{
(1) \citet{lyne93}; (2) \citet{liv2007}; (3) \citet{boyd95};  (4) \citet{walte2011}; (5) \citet{lyne96}; (6) \citet{liv2011}; (7) \citet{espinoza2011}.
}
\end{deluxetable}


\begin{deluxetable}{cccccc}
\tablecaption{\label{table:B} Ranges of  Braking Indices ($n$) Obtained with our Approach and the Actual Ones.}
\tablewidth{0pt}
\tablehead{
    \colhead{Pulsar}    & \colhead{$\left|\xi \right|$} & \colhead{$n_{estimated}$} & \colhead{$n_{estimated}$} & \colhead{n}  \\
    &  \colhead{$\times 10^{13}$} & \colhead{minimum} & \colhead{maximum} & \colhead{actual}  
}
\startdata
    Crab  & 1.5 & 1.3 & 2.5 & 2.509  \\
    PSR B1509$-$58  & 2.3 & 1.4 & 3.1 & 2.837  \\
    PSR B0540$-$69  & 4.4 & 1.3 & 2.6 & 2.140  \\
    PSR J1119$-$6127  & 7.1 & 1.5 & 3.8 & 2.684  \\
    Vela  & 11 & 0.9 & 2.3 & 1.4  \\
    PSR J1846$-$0258  & 19 & 1.8 & 3.9 & 2.16  \\
    PSR J1734$-$3333  & 34 & 0.9 & 4.6 & 0.9  \\
\enddata
\tablecomments{In the second column the  variable $\xi$, detailed in the text, is in CGS units. Columns three and four define values for the braking indices as estimated by the model; the maximum value was calculated adopting the Crab's value for $\xi$ (the smallest) while the minimum value was obtained using the value of $\xi$ for PSR J1734$-$3333 (the largest). In the column $ n \, actual $  the values of the pulsars' braking indices derived from experimental data are presented, obtained from the same references shown in the preceding table.}
\end{deluxetable}


\section{Models for Pulsars' Spin-Down}

\subsection{The Canonical Model and Typical Values}

In the canonical model the law that governs the decay of the rotation is  
\begin{equation}
\dot{\Omega} = - K \Omega ^n,
\label{eq:canon}
\end{equation}
with $n=3$ \citep{glend96} and
\begin{equation}
K \equiv   \frac{{2\mu ^2 }} {{3c^3 I}}.
\label{eq:Km}
\end{equation}
The moment of inertia of the pulsar, $I$, is of the order of $MR^2$:
\begin{equation}
I = \lambda MR^2.
\label{eq:I}
\end{equation}

The pulsar's magnetic dipole moment is $\mu$:
\begin{equation}
\mu = R^3B \sin{\alpha},
\label{eq:momDipMagn}
\end{equation}
where $R$ is the radius of the pulsar, $B$ is its surface magnetic field and $\alpha$ is the angle of inclination of the magnetic axis to the rotational axis. 

The above equations yield
\begin{equation}
|B|=\sqrt{\frac{3c^3}{2}\frac{\lambda}{\sin ^2 \alpha}\frac{M}{R^4}} \sqrt{\frac{|\dot{\Omega}|}{\Omega^n}},
\label{eq:moduloB}
\end{equation}
where we used the fact that $\dot{\Omega} <0 $. Typical values for the constants involved in this equation will be described as follows.

The constant $\lambda$ depends on the shape of the pulsar. If it is modeled as a solid ball of radius $R$ and mass $M$ then $\lambda = 2/5$; for a hollow sphere $\lambda = 2/3$. Depending on the interior structure of the star other values  may result for $\lambda$ due to the departure of the shape from spherical to ellipsoidal, for instance. In the calculations we will use the approximate value $\lambda \sim 1/2$.

Recent measurements indicate the existence of a neutron star with nearly 2 solar masses \citep{demorest10}. So far this is an isolated fact, so it still seems  reasonable to use the canonical value \citep{lat04} $M = 1.4 M_{\sun}$ in the calculations and we will assume this number.

Theoretical values for the star's radius vary from about $0.6 \times 10^6$ to $1.4 \times 10^6$cm \citep{lat04} so we will adopt the usual value $R=10^6$cm, and we will choose $\sin ^2 \alpha = 1$  so that lower limits will be implied. 

Using the above values the first square root in equation (\ref{eq:moduloB}) yields:
\begin{equation}
S \equiv \sqrt{\frac{3c^3}{2}\frac{\lambda}{\sin ^2 \alpha}\frac{M}{R^4}} = 2.3 \times 10^{20} \rm{g^{1/2}cm^{-1/2}s^{-3/2}}.
\label{eq:firstSqrt}
\end{equation}
or S =  2.3 $\times 10^{20}$   Hz$^{1/2}\,$G. This value will be considered the same for all pulsars investigated here.

Returning to equation (\ref{eq:moduloB}), since in the canonical model all pulsars have the same braking index, $n=3$,  their surface magnetic fields are different from each other as far as there are differences in the rotational behavior of the star, given by $\Omega$ and $\dot{\Omega}$. For example, among all pulsars listed in Tables \ref{table:pulsars} and \ref{table:prediction} only the X-ray pulsars present $|B|>10^{14}$G  while the others present magnetic field intensities 10 times smaller or less. These  differences have motivated the classification of the high $|B|$ pulsars as magnetars. In this work we will not focus on particular differences in $|B|$ but on general aspects of the pulsar physics. 

Since actual braking indices are not equal to 3, in the next section we investigate a modification of this model.

\subsection{The Modified Model}

We now analyze the following model for the pulsar's spin down:
\begin{equation}
\dot{\Omega} = - \bar{K}(t) \Omega ^{n_{actual}},
\label{eq:proposal}
\end{equation}
with $n_{actual}$ being the experimental value of the pulsar's braking index. The main motivation in assuming this relation is that it naturally yields $n = n_{actual}$ since it varies with each pulsar. The object $\bar{K}$ contains a number of physical characteristics of the pulsar like mass, radius, intensity of the magnetic field, perhaps internal densities, temperatures, etc.  These quantities are expected to naturally vary with time due to different processes \citep{blandRo88}. 

Inspired in the basic physics of the canonical model $\bar{K}$ is given by: 
\begin{equation}
\bar{K} = Kb^2,  
\label{eq:barK}
\end{equation}
where $K$ is given by equation (\ref{eq:Km}) and $b^2$ is a positive function of time whose unit in the CGS system is Hz$^{ 3 - n_{actual}}$. This unit (specifically, the difference in the power) suggests that $b^2$ carries information related to the correction that must be made in the canonical  model for the exact braking index be achieved, but at this point the physical content of $b$ is unknown. 

Using the same procedure as before we find an expression analogous to equation (\ref{eq:moduloB}):
\begin{equation}
\xi \equiv |B|b  =  \pm S \sqrt{\frac{|\dot{\Omega}|}{\Omega^{n_{actual}}}}.
\label{eq:weighedB}
\end{equation}
The object $\xi$ contains more physical information about the star than just the intensity of its magnetic field due to the presence of $b$. In this model $|B|$ cannot be calculated because $b$ is unknown.
In what follows we will analyze $\xi$ instead of $b$ because the right hand side of equation (\ref{eq:weighedB})  allows estimates using the constants presented previously (through the use of equation (\ref{eq:firstSqrt})) as well as the experimental values of $\Omega$, $\dot{\Omega}$ and $n_{actual}$ (or, equivalently, $\ddot{\Omega}$). It is the presence of  $n_{actual}$ in this calculation that provides more information on the pulsar's physics than the canonical model is capable of.

\section{Results}

For the pulsars that have measured values of $n$ we used equation (\ref{eq:weighedB}) as described above and calculated numerical estimates for $\left| \xi \right|$, which are listed for the different pulsars in the second column of Table \ref{table:B}, covering a range from $\left| \xi \right|_{Crab} = 1.5\times 10^{13}$ (for Crab) to $ \left| \xi \right|_{1734} = 34\times 10^{13}$ (for PSR J1734$-$3333) in units of Hz$^{(3-n_{actual})/2}\,$G. 

In order to predict the values of braking indices we deduced the following relation from equations (\ref{eq:firstSqrt}) - (\ref{eq:weighedB}):
\[
\dot{\Omega} = - \left(\frac{\xi }{S} \right)^2 \Omega  ^{n_{actual}},
\]
which implies
\begin{equation}
n_{actual} = \frac{\log \left(  \left| \dot{\Omega} \right| S^2/\xi ^2 \right)}{\log \Omega}.
\label{eq:proposal_n}
\end{equation}
An interesting fact about this relation is that if $|\xi|$ is estimated then the braking index can also be estimated only from the knowledge of the measured values of $\nu_0$ and $\dot{\nu_0}$ since $S$ is fixed due to the use of typical values of the star's characteristics.

We found a way to estimate $|\xi|$ without needing to know $\left|B\right|$ and $b$, as follows: we will assume that pulsars that have values of $\nu_0$ and $\dot{\nu}_0$ close to those of the seven pulsars of Table \ref{table:pulsars} would have a value of  $\left| \xi \right|$ within the range $\left| \xi \right|_{Crab} \leq \left| \xi \right| \leq \left| \xi \right|_{1734}$. It is reasonable to assume this as far as  $\nu_0$, $\dot{\nu}_0$ and $n_{actual}$ do not vary much and $S$ is a constant. 

Therefore, given a pulsar with $\nu_0$ and $\dot{\nu}_0$ as mentioned we use equation (\ref{eq:proposal_n}) to estimate a  range of values for that pulsar's braking index. To this end in that equation we substitute $\left| \xi \right|_{Crab}$ for the minimum value of $\left| \xi \right|$ and $\left| \xi \right|_{1734}$ for the maximum.  We tested this approach with the seven pulsars and in Table \ref{table:B} we present the results.

We also did this exercise with four radio pulsars and two anomalous X-ray pulsars believed to be magnetars. In Table \ref{table:prediction} we present the results for these objects which do not have known braking indices so far. This approach predicts that the actual values of these pulsars' braking indices should lie within the range between  $n_{minimum}$ and $n_{maximum}$ in the respective lines of that table.

\section{Discussion}

For the family of pulsars that we are analyzing the spin frequency and its first time derivative are in the range 
$0.8 Hz \leq \nu_0 \leq 31$ Hz and $-3 \leq \dot{\nu}_0 \leq -0.01$ ($\times 10^{-10}$ s$^{-2}$) (see Table \ref{table:B}). According to the catalog \citet{pulsarCatalog12} there are tens of pulsar within these ranges. Using this family
 the quantity $|\xi|$ is expected to lie within a range as well, given by $1.5 \leq |\xi|\leq 34$($\times 10^{13}$ Hz$^{(3-n_{actual})/2}\,$G)(see Table \ref{table:prediction}).

The object $\xi$ may vary with time since it depends, for instance, on $\nu_0$ and $\dot{\nu}_0$ as well as on $B$ (see equation (\ref{eq:weighedB})). For our study, however, we did not need values for the magnetic field (a topic that \citet{blandRo88}  explore, for instance). The values of $B$ that the pulsars may have will depend on the object $b$ which we did not need to address in the present analysis and is open for now. This quantity, like $\xi$, is expected to contain information on particularities of the star besides its spin behavior, like internal structure, stresses, interaction with the environment etc. 

In Table \ref{table:B} the ranges from ``$n_{estimated}\, minimum$" to ``$n_{estimated}\,maximum$" for all those pulsars do contain the respective values of their actual braking indices. There are some ``$n_{estimated}\,maximum$" larger than the canonical value ($n=3$), indicating that the Crab's value for $|\xi|$ is too small for such pulsars.

In Table \ref{table:prediction} we chose those pulsars as examples for the following reasons: (i) PSR B2334+61,  PSR J1418$-$6058 and PSR J1124$-$5916 have relatively high values for $\dot{\nu}_0$ thus being perhaps good candidates for the experimental determination of $\ddot{\nu}_0$; (ii) PSR  J1509$-$5850 has a limiting value for $\dot{\nu}$; and (iii) 1E 1547.0$-$5408 and 1E 1841$-$045 are X-ray pulsars and their frequencies or frequency derivatives are somewhat departed from the ranges of interest. 

The first four  pulsars in that table show braking index ranges similar to those in Table \ref{table:B}. In particular, PSR J1509$-$5850 may have a braking index significantly lower than 3.

The X-ray pulsars show very different values for $n_{estimated}\,maximum$ which indicate that they might have important physical differences compared to the pulsars in Table \ref{table:B}. It seems that 1E 1547.0$-$5408 has braking index not far from 3.



\begin{deluxetable}{cccccc}
\tablecaption{\label{table:prediction} Values for the braking indices of pulsars calculated using the proposed model. The maximum value was calculated adopting the Crab's value for $\xi$ while the minimum value was obtained using the value of $\xi$ for PSR J1734$-$3333. The sources for the values of $\nu_0$ and $\dot{\nu}_0$ are listed in the last column. Pulsar 1E 1841$-$045  is also known as PSR J1841$-$0456 while 1E 1547.0$-$5408  is the same as PSR J1550$-$5418. }
\tablewidth{0pt}
\tablehead{
   \colhead{Pulsar} & \colhead{$\nu_0$ (Hz)}  & \colhead{$\dot \nu_0$ ($\times 10^{-10} s^{-2}$)} &  \colhead{$n_{minimum}$} & \colhead{$n_{maximum}$} & \colhead{Refs} 
   } 
\startdata
        PSR B2334+61 & 2.02 & $-$0.0079 & 0.3 & 2.7  & 1 \\
    PSR  J1509$-$5850 & 11.2
 & $-$0.0116 & 0.3 & 1.7 & 2  \\
    PSR J1418$-$6058 & 9.04 & $-$0.1385 & 0.9 & 2.4 & 3 \\
    PSR J1124$-$5916 & 7.38 & $-$0.4100 & 1.2 & 2.8 & 4  \\
    X-ray pulsars: & & & & & \\
    1E 1547.0$-$5408 &  0.483 & $-$0.0541 & 2.4 & 8.0 & 5 \\
    1E 1841$-$045 & 0.848 & $-$3.216 & 1.2 & $-$9.7 & 6,7
\enddata
\tablerefs{
(1) \citet{yuan2010}; (2) \citet{kram2003};  (3) \citet{abdo09}; (4) \citet{cam02}; (5) \citet{cam07}; (6) \citet{vasis97}; (7) \citet{kuip06}.
}

\end{deluxetable}


\section{Concluding remarks}

We analyzed a modification of the canonical model for pulsars' spin down that introduces in the equation parameters ($\xi$ or $b$) with physical content of the star. We investigated particularly the parameter $\xi$ which is sensitive to the star's braking index and thus to its physical characteristics.

Using in the analyzed model the values for the braking index of seven pulsars, obtained so far from experimental data,  we found ranges for these pulsars' braking indices in which all actual braking indices laid. We calculated $|\xi|$  for these pulsars and used them to predict ranges of braking indices for other pulsars, shown in Table \ref{table:prediction}. These results were possible assuming that all pulsars have the same (typical) values for some physical characteristics, like mass, radius etc.

In order to improve the model it is necessary to find physical details about $\xi$; this would help limit it to better values. It is expected to provide information on the pulsar's magnetic field as the definition shows, equation (\ref{eq:weighedB}), but contains also a contribution from the unknown object $b$. We are working on the proposal of a physical context that would yield the parameters $\xi$ and $b$ and in which mass, radius and other parameters are allowed to vary.

\acknowledgments

NSM acknowledges Dr. Jose A. S. de Lima for introducing her to the braking index problem. TAM acknowledges the Brazilian institution UNIFESP for financial support. NSM has a fellowship from the Brazilian federal funding agency CNPq (grant 309295/2009-2).


\begin{thebibliography}{}

\bibitem[Abdo et al.(2009)]{abdo09} Abdo, A. A., Ackermann, M., Ajello, M., et al. 2009, Science, 325, 840 

\bibitem[ATNF/CSIRO(2012)]{pulsarCatalog12} ATNF/CSIRO. 2012, ATNF Pulsar Catalogue, Web Interface to Database. http://www.atnf.csiro.au/research/pulsar/psrcat/. [Also: Manchester, R. N., Hobbs, G. B., Teoh, A. \& Hobbs, M. 2005,  \apj, 129, 1993/


\bibitem[Blandford \& Romani(1988)]{blandRo88} Blandford, R. D. \& Romani R. W. 1988, \mnras, 234, 57P


\bibitem[Boyd et al.(1995)]{boyd95} Boyd, P.T., van Citters, G. W., Dolan, J. F., et al. 1995, \apj, 448, 365

\bibitem[Camilo et al.(2002)]{cam02} Camilo, F.,Manchester, R. N., Gaensler, B. M., Lorimer, D. R., \& Sarkissian,
J. 2002, \apj, 567, L71 

\bibitem[Camilo et al.(2007)]{cam07} Camilo, F., Ransom, S. M.,  Halpern, J. P. \&  Reynolds, J. 2007, \apj, 666, L93 

\bibitem[Contopoulos \& Spitkovsky(2006)]{contop06} Contopoulos, I. \& Spitkovsky, A. 2006, \apj, 643, 1139

\bibitem[Demorest et al.(2010)]{demorest10} Demorest, P. B., Pennucci, T., Ransom, S. M., Roberts, M. S. E., \& Hessels, J.
W. T. 2010, Nature, 467, 1081

\bibitem[Espinoza et al.(2011)]{espinoza2011} Espinoza, C. M.,al.Lyne, A. G., Kramer, M., Manchester, R. N., \& Kaspi, V. M. 2011, ApJ, 741, L13

\bibitem[Glendenning(1996)]{glend96} Glendenning, N. K. 2000, Compact Stars (2nd ed.; New York: Springer)

\bibitem[Kaspi et al.(1994)]{kaspi94} Kaspi, V. M., Manchester, R. N., Siegman, B., Johnston, S., \& Lyne, A. G. 1994, \apj,  422, L83

\bibitem[Kramer et al.(2003)]{kram2003} Kramer, M.,Kramer, M., Bell, J. F., Manchester, R. N., et al. 2003, \mnras,  342, 1299


\bibitem[Kuiper et al. (2006)]{kuip06} Kuiper, L., Hermsen, W., den Hartog, P. R. \& Collmar,  W. 2006, \apj, 645, 556

\bibitem[Lattimer \& Prakash(2004)]{lat04} Lattimer, J. M. \&  Prakash,  M. 2004, Science, 304, 536

\bibitem[Livingstone et al.(2007)]{liv2007} Livingstone, M. A., Kaspi, V. M., Gavriil, F. P., et al. 2007, \apss, 308, 317

\bibitem[Livingstone et al.(2011)]{liv2011} Livingstone, M. A., Ng, C.-Y., Kaspi, V. M., Gavriil, F. P., \& Gotthelf, E. V. 2011, \apj, 730, 66

\bibitem[Lorimer(2008)]{lorimer08} Lorimer, D. R. 2008, Living Rev. Relativity 11, 8, http://www.livingreviews.org/lrr-2008-8 

\bibitem[Lyne, Pritchard \& Graham Smith(1993)]{lyne93} Lyne, A. G., Pritchard, R. S. \& Graham Smith, F. 1993, \mnras, 265, 1003

\bibitem[Lyne et al.(1996)]{lyne96} Lyne, A. G. , Pritchard,  R. S., Graham Smith,  F. \& Camilo, F. 1996, Nature, 381, 497


\bibitem[Manchester \& Taylor(1977)]{man77} Manchester, R. N.  \&  Taylor, J. H. 1977, Pulsars (San Francisco, CA: W. H. Freeman \& Co.)

\bibitem[Melatos(1997)]{mela97} Melatos, A. 1997, \mnras, 288, 1049

\bibitem[Ostriker \& Gunn(1969)]{osg} Ostriker, J. P.  \& Gunn, J. E. 1969, \apj, 157, 1395

\bibitem[Vasisht \& Gotthelf(1997)]{vasis97} Vasisht, G. \& Gotthelf, E. V. 1997, \apj, 486, L129

\bibitem[Waltevrede, Johnston \& Espinoza(2011)]{walte2011} Waltevrede, P.,  Johnston, S. \&  Espinoza, C. M. 2011, \mnras, 411, 1917

\bibitem[Yuan et al.(2010)]{yuan2010} Yuan, J. P., Manchester, R. N., Wang, N., et al. 2010, \apj,  719, L111

\end{thebibliography}
\end{document}